\documentclass[prb,aps,twocolumn,floatfix]{revtex4-1}
\usepackage{amssymb,graphicx,color,amsmath,bm,subfigure}
\usepackage[bookmarks, colorlinks=true, breaklinks]{hyperref}
\usepackage[none]{hyphenat}
\hypersetup{linkcolor=blue,citecolor=blue,filecolor=black,urlcolor=blue}

\newcommand{\bra}{\left< }
\newcommand{\ket}{\right>}
\newcommand{\ovl}[1]{\overline{#1}}

\begin{document}
\title{Paraelectric and Ferroelectric States in a Model for Relaxor Ferroelectrics}
\author{G. G. Guzm\'{a}n-Verri$^{1}$, P. B. Littlewood$^{2,3}$, and C. M. Varma$^{4}$}
\affiliation{$^1$Materials Science Division, Argonne National Laboratory, Argonne, IL 60439 \\
$^2$Physical Sciences and Engineering, Argonne National Laboratory, Argonne, IL 60439 \\
$^3$James Franck Institute, University of Chicago, Chicago, IL 60637 \\
$^4$Department of Physics and Astronomy, University of California, Riverside, CA 92521}

\begin{abstract}
We study the free energy landscape of a minimal model for relaxor ferroelectrics.
Using a variational method which includes leading correlations beyond the mean-field approximation as well as disorder averaging at the level of a simple replica theory, we find 
metastable paraelectric states with a stability region that extends to zero temperature.
The free energy of such states exhibits an essential singularity for weak compositional disorder
pointing to their necessary occurrence. Ferroelectric states appear as local minima in the free energy at high temperatures and become stable below a coexistence temperature $T_c$. 
We calculate the phase diagram in the electric field-temperature plane and find a 
coexistence line of the polar and non-polar phases which ends at a critical point.
First-order phase transitions are induced for fields sufficiently large to cross
the region of stability of the metastable paraelectric phase. 
These polar and non-polar states have distinct structure factors from those of conventional ferroelectrics. 
We use this theoretical framework to compare and to gain physical understanding of various experimental results in typical relaxors. 
\end{abstract}
\date{\today}
\maketitle

\section{Introduction}

The unusual linear and nonlinear dielectric response of relaxor ferroelectrics make them both technologically important and scientifically remarkable.~\cite{Cowley2011a, Bokov2006a, Kleeman2006a, Samara2003a, Cross1987a}
Typical relaxors such as PbMn$_{1/3}$Nb$_{2/3}$O$_3$~(PMN) and PbZn$_{1/3}$Nb$_{2/3}$O$_3$~(PZN) show extended regions of fluctuations (often called diffuse phase transitions) 
with several special energy scales:
for temperatures above the so-called Burns temperature $T_B$~\cite{Burns1983} their dielectric constant follows the Curie-Weiss law with a 
characteristic Curie-Weiss temperature $T_{CW}$.~\cite{Viehland1992a} Below $T_B$, it deviates from the Curie-Weiss behavior and reaches a broad, frequency dependent maximum at a temperature $T_{max}(\omega)$ without any signature of a global broken symmetry.~\cite{Smolenskii1970a, Bovtun2004a, Hlinka2006a, Bokov2012a} $T_{max}(\omega)$ follows the Vogel-Fulcher law.~\cite{Glazounov1998a} At low temperatures,
no macroscopic structural changes are observed~\cite{Bonneau1989a, deNathan1991a} unless large enough electric fields are applied, which reveal an additional energy scale $T_c$.~\cite{Schmidt1980a, Kutnjak2006a, Zhao2007a, Bing2005a} 
Very significantly, neutron scattering experiments observe the onset of 
elastic diffuse scattering at a temperature $T^*$
with unusual temperature dependence~\cite{Hiraka2004a, Gvasaliya2005a, Rotary2008a, Gehring2009a} and non-Lorentzian line shapes.~\cite{Gvasaliya2005a}
It is found that $T^* \simeq T_{CW}$ within experimental uncertainty.~\cite{Gehring2009a, Gehring2012a}
Under static conditions, 
$T_c$ and $T^*$ are the only 
temperature scales observed in elastic X-ray 
and neutron scattering experiments.~\cite{Stock2007a, Gvasaliya2005a, Gehring2009a}  
Solid solutions of relaxors with conventional ferroelectrics such as PbMn$_{1/3}$Nb$_{2/3}$O$_3$-PbTiO$_3$~(PMN-PT) and PbZn$_{1/3}$Nb$_{2/3}$O$_3$-PbTiO$_3$~(PZN-PT) also show relaxor behavior for small PT content and well defined ferroelectric transitions for sufficiently large PT.~\cite{Colla1998a, Shuvaeva2005a, La-Orauttapong2002a}
Transitions of structurally distinct ferroelectric phases across a morphotropic phase boundary occur
for intermediate PT concentrations.~\cite{Chois1996a, Noblanc1996a, Kelly1997a, Park1997a, Ye2000a}

Though relaxors were first synthesized more than 50 years ago,~\cite{Roy1954a, Smolenskii1958a} 
and several models have been proposed to describe their dielectric behavior,~\cite{Blinc1999a, Glinchuk2004a, Burton2006a, Tinte2006a, Akbarzadeh2012a, Takenaka2013a} 
several aspects of their properties are not well understood. 
A common problem is that it is difficult to assign the  
parameters of a model to their characteristic temperatures in a universal fashion.~\cite{Cowley2011a}
An additional difficulty has been to identify the ground state or glassy metastable states of 
relaxors due to their skin effect. 
For PMN, a cubic-to-rhombohedral distortion is observed at about $T_c$ in 
the near-surface region~(the skin) while the bulk remains cubic down to low temperatures.~\cite{Stock2007a, Conlon2004a}
The skin is macroscopically large~(a few tens of micrometers)
and thus it is not clear whether the skin or the bulk 
is in a thermodynamic stable state. 
Similar skin effects have been observed in PZN,~\cite{Xu2003a} 
PMN-PTO and PZN-PTO.~\cite{Gehring2004a, Xu2006a}

Compositional disorder is essential to observe relaxor behavior.~\cite{Randall1990a}
In the heterovalent relaxor PMN, for instance, disorder arises from the different charge valencies and atomic radii 
of Mg$^{+2}$ and Nb$^{+5}$ on the octahedrally coordinated site.~\cite{Cowley2011a}
Such disorder leads to (i) quenched random electric fields~\cite{Westphal1992a} and (ii) quenched random bonds.~\cite{Blinc1999a}
It is expected that (i) introduces effects similar to those in
magnets with quenched random magnetic fields~\cite{Imry1975a, Young1991a}  while (ii) is the classic ingredient
 of spin glasses if the bonds are frustrated.~\cite{FischerBook} If they are not, this is the random local transition temperature model in Landau theory of phase transitions.~\cite{Dotsenko1995a}
Quite generally, it is understood in the theory of phase transitions that 
fields which couple linearly to the order parameter affect the properties much more strongly than random bonds
which couple to the order parameter quadratically.
This point has been appreciated in connection with relaxors.~\cite{Cowley2011a} 
Therefore,
the effects of (i) must be understood before including (ii).
Experiments on the non-ergodic
behavior of relaxors,~\cite{Vakhrushev1989a, Westphal1992a, Zhao2007a} their skin effect~\cite{Conlon2004a, Xu2003a, Gehring2004a, Xu2006a, Stock2007a} and
their structure factors~\cite{Gvasaliya2005a, Gehring2009a} provide support for this view as similar
effects have been observed in magnets with quenched random fields.~\cite{neutrons_RFIM, Hill1991a}

The purpose of this paper is to study the effects on quenched 
electric random fields in a simple displacive model 
with which conventional ferroelectrics were first understood.~\cite{Cochran1959a, Anderson1960a} This model was recently used in Ref.~[\onlinecite{Guzman2012a}] to show that the broad region of fluctuations~(a hallmark of relaxors)
is the result of dipolar interactions~(already present in the displacive model) acting together with compositional disorder. 
This point can be appreciated from the fact 
noted by Onsager,~\cite{Onsager1936a} that due to intrinsic fluctuations a model with dipolar interactions alone has no phase transition down to the lowest temperature.  
Similar models with further extensions have been studied earlier, however, they have almost exclusively been studied by numerical methods.~\cite{Burton2006a, Tinte2006a} Analytic solutions often help in finding general features of the solution, therefore 
we consider the minimal model of Ref.~[\onlinecite{Guzman2012a}]. 
Though by no means exhaustive, this analysis allows
an analytic account of microscopic aspects of the physics of relaxors. 
In this paper we introduce a method more general than that used in Ref.~[\onlinecite{Guzman2012a}] 
and also study the effects of static applied electric fields. 

The essential physical points in the simplest necessary solution of the model are to formulate an approximation which considers thermal and quantum fluctuations at least at the level of the Onsager approximation,~\cite{Onsager1936a} and treats compositional disorder at least at the level
of a replica theory.~\cite{DeDominicis_book} 
Here, we do so with a variational method which extends the self-consistent phonon approximation~\cite{Jones_book} 
 to incorporate disorder.~\cite{Narimanov2000}
This variational method leads to the self-consistent equations of the
earlier approximation, but also allows  us
to explore the energy landscape and the competition 
between states with and without spontaneous polarization.
We show that there are metastable paraelectric states in the free energy
that persist down to zero temperature. 
Within our approximation,
the free energy of the disordered state exhibits an
essential singularity for weak disorder.
Ferroelectric states appear as local minima in the free energy 
above a coexistence temperature $T_c$ and
become stable below it. 
We calculate the electric field-temperature $(E-T)$ phase diagram for moderate disorder and find a 
coexistence line of the polar and non-polar phases 
which ends at a critical point. 
First-order phase transitions are induced for fields sufficiently strong
to cross the stability line of the metastable paraelectric phase. 
These ordered and disordered states are unusual, as their
structure factors differ from those of conventional ferroelectrics. 

This paper is organized as follows: in Sections~\ref{sec:Model_Hamiltonian} 
and~\ref{sec:Variational_Solution} we present our model Hamiltonian and
variational solution, respectively; the results and discussion are presented 
in Section \ref{sec:RD}; and a comparison to experiments is provided in Section~\ref{sec:comparison}. 
A summary and conclusions are given Section~\ref{sec:Conclusions}. 

\section{Model Hamiltonian}
\label{sec:Model_Hamiltonian}

We consider the model for relaxor ferroelectrics of Ref.~[\onlinecite{Guzman2012a}]
and add an applied field.
We focus on the relevant transverse optic mode configuration coordinate  $u_i$ of the ions in the unit cell $i$ along the polar axis (chosen to be the $z$-axis).  $u_i$ experiences a local random field $h_i$ with probability $P( h_1, h_2,...)$ due to the compositional disorder. The model Hamiltonian is
\begin{align}
\label{eq:Hamiltonian_RFE}
H&=\sum_{i}\left[\frac{\Pi_i^2}{2M}+V(u_i)\right]-\frac{1}{2}\sum_{i,j}v_{ij}u_i u_j \nonumber\\ 
&~~~~~~~~~~~~~~~~~~~~~~~~~~~~~~~~~~
-\sum_i h_i u_{i} - E_0 \sum_i u_i,
\end{align}
where $\Pi_i$ is the momentum conjugate to $u_i$, $M$ is an effective mass, and  $E_0$ is
a static applied electric field.
We assume the $h_i$'s are independent random variables with Gaussian probability distribution with zero mean and variance $\Delta^2$,
\begin{align}
P( h_1, h_2,...) = \prod_{i} \frac{1}{\sqrt{ 2 \pi \Delta^2 } } \, e^{ - \frac{1}{2} \frac{ h_i^2 }{ \Delta^2 } }.
\end{align}

$V(u_i)$ is an anharmonic potential,
\begin{equation}
\label{eq:anharmonic_potential_RFE}
V(u_i)=\frac{\kappa}{2} u_i^2 + \frac{\gamma}{4} u_i^4,
\end{equation}
where $\kappa,~\gamma$ are positive constants.
$v_{ij}$ is the dipole interaction,
\begin{equation}
\label{eq:vij}
v_{ij}/{e^*}^2=
\begin{cases}
3 \frac{(Z_{i}-Z_{j})^2}{|{\bm R}_i-{\bm R}_j|^5} -\frac{1}{|{\bm R}_i-{\bm R}_j|^3}, & {\bm R}_i \neq {\bm R}_j\\
0, & {\bm R}_i = {\bm R}_j,
\end{cases}
\end{equation}
where $e^*$ is the effective charge and $Z_i$ is the $z$-component of ${\bm R}_i$.

The Hamiltonian of Eq.~(\ref{eq:Hamiltonian_RFE}) presents long ranged (anisotropic)
dipolar interactions, compositional disorder, and anharmonicity. 
We  do not consider cubic symmetry, coupling to strain fields and
disorder in the bonds $v_{ij}$ expected in relaxors.~\cite{Cowley2011a, Blinc1999a, Burton2006a, Tinte2006a} 
As we stated above, our purpose is to study the effects
of quenched random fields alone in a displacive model for ferroelectrics.

\section{Variational Solution}
\label{sec:Variational_Solution}

In this section, we present a variational framework
to study the statistical mechanics of the problem
posed by the Hamiltonian~(\ref{eq:Hamiltonian_RFE}).

We consider a trial pair-probability distribution,
\begin{equation}
\label{eq:rho_tr}
\rho^{tr} = \frac{1}{Z^{tr}}e^{-\beta H^{tr}},
\end{equation}
where $H^{tr}$ is the Hamiltonian of  coupled displaced harmonic oscillators in a random field,
\begin{equation}
\label{eq:Htr}
H^{tr} = \sum_i \frac{\Pi_i^2}{2M} + \frac{1}{2}\sum_{i,j} ( u_i - p ) G_{i-j}  ( u_j - p ) - \sum_{i} h_i u_i,
\end{equation}
and $Z^{tr}$ its normalization,
\begin{align}
\label{eq:Ztr}
Z^{tr} = \mbox{Tr} e^{- \beta H^{tr} } 
& = \left( \prod_{\bm q}  \left[ 2 \sinh \left( \frac{\beta \hbar \Omega_{\bm q} }{2} \right) \right]^{-1} \right) 
\nonumber \\
&~~~~~~~~~~~
\times
\left( \prod_{i,j} e^{\frac{1}{2} \beta h_i  G_{i-j}^{-1}  h_j  + \beta h_i p } \right).
\end{align}
Here, $p$ is an uniform order parameter: it is the displacement coordinate averaged over thermal disorder~($ \bra \dots \ket $) first,
and then over compositional disorder~($ \overline{ \bra  \dots \ket } $),
\begin{align}
\label{eq:order_parameter}
p = \overline{ \bra u_i \ket }= \int_{-\infty}^\infty dh_1 dh_2 \cdots P(h_1,h_2,...) \, \mbox{Tr} \, \rho^{tr} \, u_i.
\end{align}
In the standard variational scheme of the self-consistent phonon approximation, 
the Fourier transform of the function $G_{i-j}$ is the frequency of the transverse optic mode
$ M \Omega_{\bm q}^2 =  \sum_{i,j} G_{i-j} e^{ i {\bm q} \cdot ( {\bm R}_i - {\bm R}_j ) } $, at wavevector ${\bm q}$.~\cite{Jones_book} 
We define $  G_{i-j}^{-1} = \left( 1 / N \right) \sum_{\bm q} (M \Omega_{\bm q}^2 )^{-1}  e^{ -i {\bm q} \cdot ( {\bm R}_i - {\bm R}_j ) } $ where
the summations over ${\bm q}$ extend over the first Brillouin zone.
$p$ and $\Omega_{\bm q}  $ are variational parameters and are determined by minimization of the free energy.

Using Eqs.~(\ref{eq:Hamiltonian_RFE})-(\ref{eq:order_parameter}), we calculate
the free energy, $ \overline{F} =   \overline{ \bra H \ket } + T  \overline{ \bra k_B \ln \rho^{tr} \ket } $.
The result is given as follows,
\begin{widetext}
\begin{align}
\label{eq:free_energy}
\frac{\ovl{F}}{N} &= 
\frac{ \kappa }{ 2 } \left[ p^2 + \eta + \Delta^2 \psi \right] + 
\frac{\gamma}{4}\left[ p^4 + 6 p^2 \Delta^2 \psi + 3 \Delta^4 \psi^2  
+ 6 \eta \left\{ p^2 + \Delta^2 \psi  \right\} + 3 \eta^2 \right] \nonumber \\
&-
\frac{1}{2} \frac{1}{N} \sum_{\bm q} v_{\bm q} \frac{\hbar}{2 M \Omega_{\bm q} } \coth\left( \frac{\beta \hbar \Omega_{\bm q}}{2} \right)
-\frac{1}{2}v_0^\perp p^2
-\frac{ 1}{2} \frac{1}{N} \sum_{\bm q} v_{\bm q} \frac{\Delta^2}{ \left( M \Omega_{\bm q}^2 \right)^2 } \nonumber \\
&- 
 \frac{1}{N} \sum_{\bm q}\frac{\Delta^2}{ M \Omega_{\bm q}^2  } 
- E_0 p  \\
&- \frac{1}{4} \frac{1}{N} \sum_{\bm q} \hbar \Omega_{\bm q} \coth\left( \frac{\beta \hbar \Omega_{\bm q}}{2} \right) + \frac{k_B T}{N} \sum_{\bm q} \ln \left[ 2 \sinh \left( \frac{\beta \hbar \Omega_{\bm q} }{2} \right) \right], \nonumber
\end{align}
\end{widetext}
where $ v_{\bm q} / ( n {e^*}^2 ) = 1/ ( n {e^*}^2 )  \sum_{i,j}v_{ij} e^{ i {\bm q} \cdot ( {\bm R}_i - {\bm R}_j ) }= \frac{4\pi}{3}\left(1-3 \frac{q_z^2}{|{\bm q}|^2}\right)-\zeta |{\bm q} a|^2 + 3\zeta (q_z a)^2 $ is the Fourier component of the dipole interaction $v_{ij}$ for cubic lattices in the long-wavelength limit; $\zeta$ is a dimensionless coefficient that depends on the structure of the lattice;~\cite{Aharony1973a} $a$ is the lattice constant; and $v_{0}^{\perp}~ = 4\pi n {e^*}^2 / 3 $
the ${\bm q}=0$ component of $v_{\bm q}$ in the direction transverse to the polar axis~($v_{\bm q}$ is non-analytic for ${\bm q} \to 0$). 
$\eta$ are mean squared fluctuations averaged over compositional disorder,
\begin{align}
\label{eq:eta}
\eta 
&= \overline{ \bra \left( u_i - \bra u_i \ket \right)^2 \ket } 
= \frac{1}{N} \sum_{\bm q} \frac{\hbar}{2 M \Omega_{\bm q} } \coth\left( \frac{\beta \hbar \Omega_{\bm q}}{2} \right),
\end{align}
and $\psi$ is defined as follows,
\begin{align}
\label{eq:xi}
\psi \equiv  \frac{1}{N}\sum_{\bm q} \frac{1}{ \left( M \Omega_{\bm q}^2 \right)^2 }.
\end{align}
We have ignored terms independent of $p$ and $ \Omega_{\bm q} $ in Eq.~(\ref{eq:free_energy}).

Minimization of the free energy with respect to $p$ and $ \Omega_{\bm q} $ gives the result,
\begin{subequations}
\label{eq:euler-lagrange}
\begin{align}
\label{eq:euler-lagrange_1}
E_0 &= \left[ M ( \Omega_0^\perp )^2 - 2 \gamma p^2 \right]p  \\
M \Omega_{\bm q}^2 &= M ( \Omega_0^\perp )^2 + \left( v_0^\perp - v_{\bm q} \right) \\
\label{eq:euler-lagrange_3}
M ( \Omega_0^\perp )^2  &=  \kappa + 3 \gamma \left[ \eta + \Delta^2 \psi + p^2 \right] - v_0^\perp.
\end{align}
\end{subequations}
Equations~(\ref{eq:euler-lagrange_1})-(\ref{eq:euler-lagrange_3}) together with Eqs.~(\ref{eq:eta}) and~(\ref{eq:xi}) are self-consistent equations that determine
the temperature dependence of the zone-center soft mode $\Omega_0^\perp$ and the order parameter $p$. The temperature dependence of $\Omega_{\bm q}$ is determined 
through $\Omega_0^\perp$.
For $E_0=0$ and $p=0$ they correspond to those derived in Ref.~[\onlinecite{Guzman2012a}].
For no disorder~($\Delta=0$), they correspond to those of the self-consistent phonon approximation for pure ferroelectrics.

We now compute the structure factor $S_{\bm q}$. 
The structure factor is obtained from the Fourier transform of the correlation functions
$ \overline{ \bra u_i u_j \ket }$. With the help of Eqs.~(\ref{eq:rho_tr})-(\ref{eq:Ztr}), we obtain the following result,
\begin{align}
\label{eq:Sq}
S_{\bm q} = p^2 \delta_{\bm q} + \frac{\hbar}{2 M \Omega_{\bm q} } \coth\left( \frac{\beta \hbar \Omega_{\bm q}}{2} \right) + \frac{\Delta^2}{ \left( M \Omega_{\bm q}^2 \right)^2 },
\end{align}
where $p,~\Omega_{\bm q} $ are given by Eq.~(\ref{eq:euler-lagrange}).
This expression corresponds to the structure factor derived in Ref.~[\onlinecite{Guzman2012a}]
with an additional contribution from the order parameter $p$.
We identify the correlation length $\xi$ from the structure factor of the pure system:
in the classical limit and for no compositional disorder, we
recover the structure factor for conventional ferroelectrics, 
$S_{\bm q} = p^2 \delta_{\bm q}  + k_B T / \left( M \Omega_{\bm q}^2 \right) $,
with $ M \Omega_{\bm q}^2 = \frac{ v_0^\perp \zeta a^2 } { 4\pi / 3 }  \left( \xi^{-2} +  \left| {\bm q} \right|^2 \right)$ for wavevectors
${\bm q}$  in the $x$ and $y$ directions. We recognize $\xi$ as the correlation length 
\begin{align}
\label{eq:correlation_length}
\xi / a = \sqrt{ \frac{ \zeta / (4\pi / 3)  }{ M ( \Omega_0^\perp )^2 / v_0^\perp  } },
\end{align}
which diverges at the onset of the ferroelectric transition for the pure system.

\section{Results and Discussion}
\label{sec:RD}

\subsection{No Applied Electric Field, $E_0=0$.}

\begin{figure}[h!]
  \begin{center}
      {\includegraphics[scale=0.60]{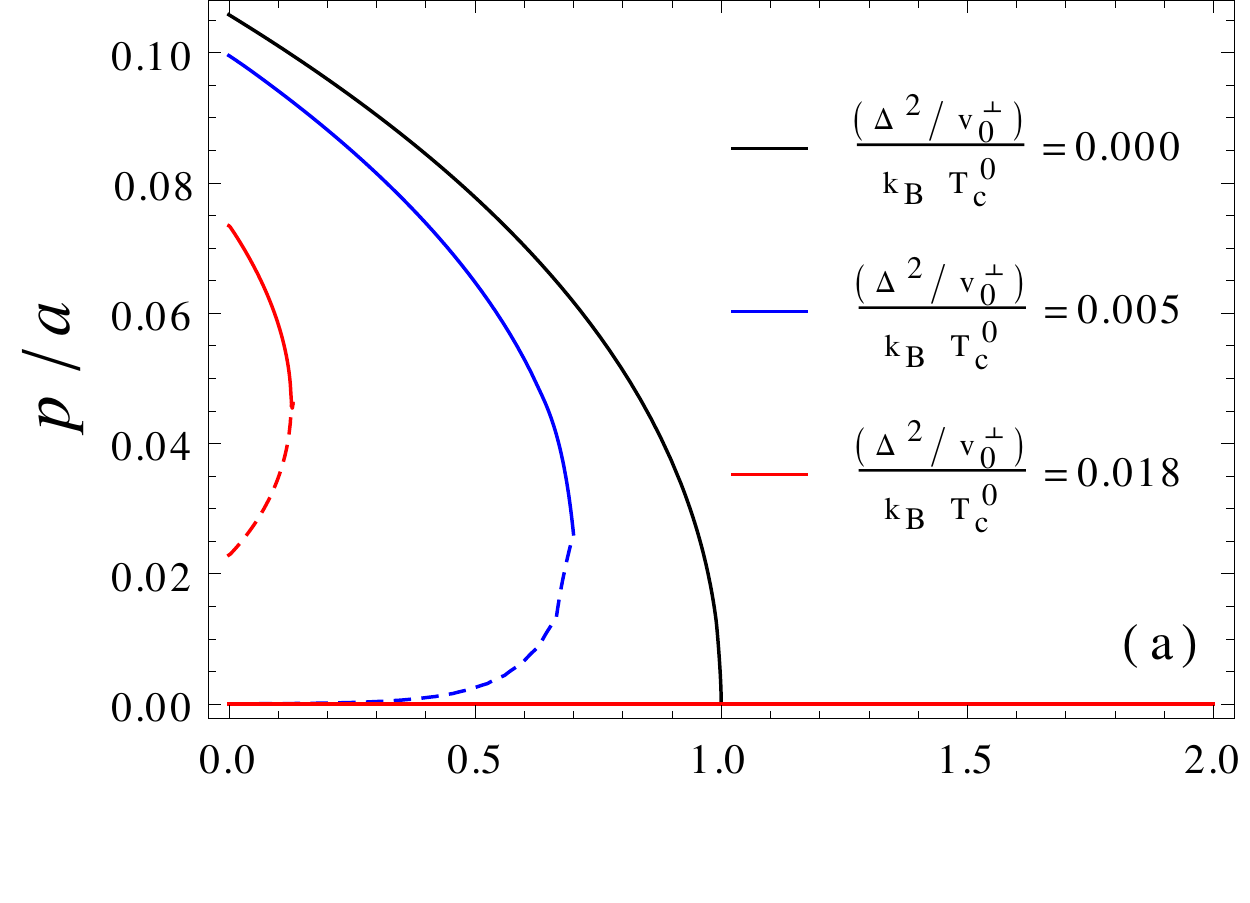}} \\ \vspace{-0.5cm}
      {\includegraphics[scale=0.60]{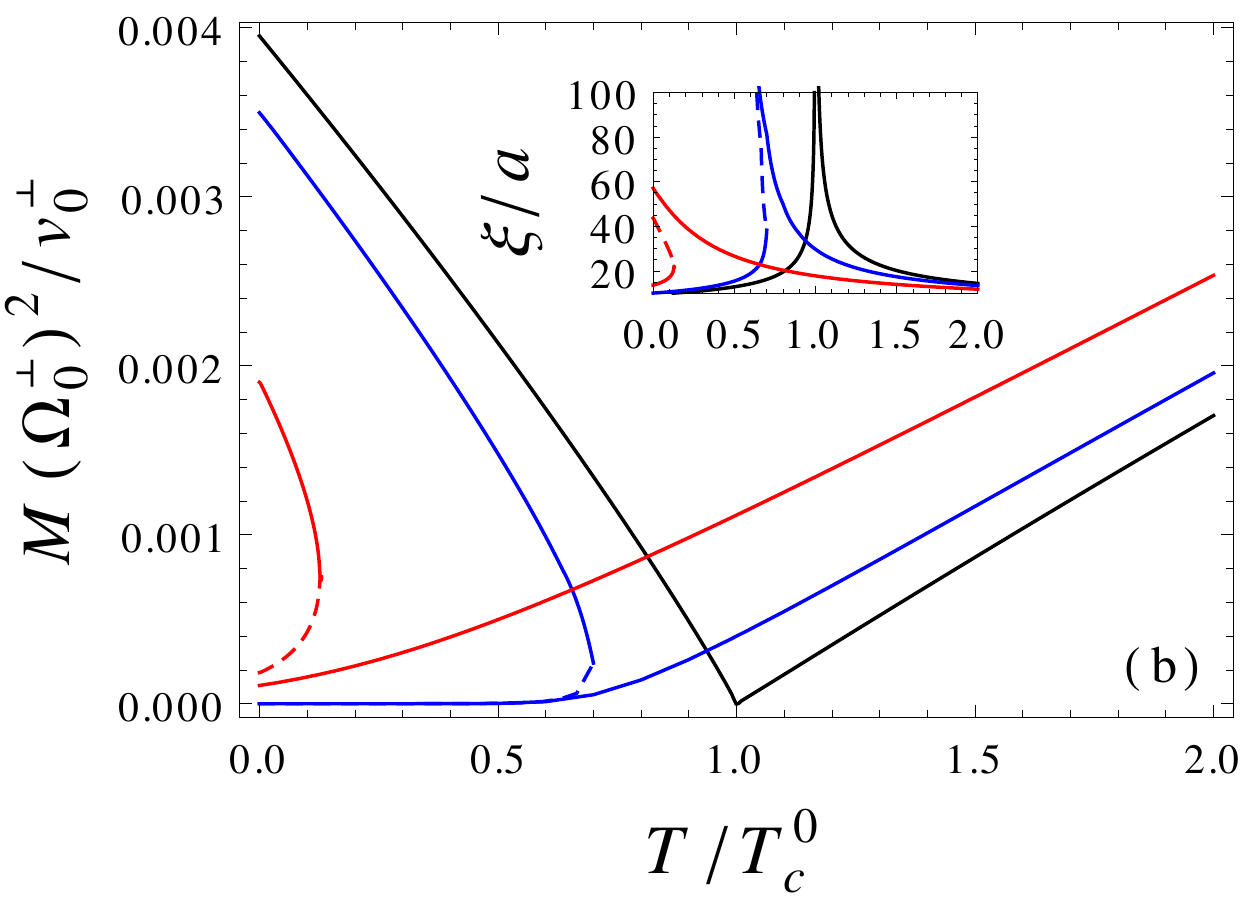}} 
  \end{center}\vspace{-0.65cm}
  \caption{(Color online) Temperature dependence of (a) the order parameter $p$ and (b) the zone-center soft mode frequency $\Omega_0^\perp$.
Inset: Temperature dependence of the correlation length $\xi$.
There is an essential singularity for small compositional disorder.
Solid lines correspond to stable and metastable states.
Dashed lines correspond to a saddle point in the free energy. 
Here, $ \hbar / ( M v_0^\perp a^4 )^{1/2} = 4.5 \times 10^{-6} $,
$(v_0^\perp - \kappa) /v_0^\perp = 1.98 \times 10^{-3}, \gamma a^2 / v_0^\perp  = 1.98 \times 10^{-1}, k_B T_c^0/ (v_0^\perp a^2) = 5.70 \times 10^{-4} $ 
These model parameters were obtained from fits to transition temperature, Curie-Weiss constant, polarization, and phonon dispersion 
of PT at $\Delta=0$.~\cite{Glass1970a, Shirane1970a}}
\label{fig:omega_p_small}
\end{figure}

Figures~\ref{fig:omega_p_small}~(a)-(b) show the temperature dependence of the order parameter and the 
the zone-center transverse optic frequency obtained from Eq.~(\ref{eq:euler-lagrange}) for $E_0=0$.
For no disorder, we obtain the well-known results of the self-consistent phonon approximation: a paraelectric-to-ferroelectric second order phase transition at a
critical temperature $T_c^0$ with a transverse optic phonon frequency that softens as  $( \Omega_0^\perp )^2 \propto |T-T_c^0|$, with logarithmic corrections.~\cite{Lines1977a}
For finite disorder, ordered states with opposite polarization appear below a superheating temperature $T_1 < T_c^0$~(only the state with positive polarization is shown). 
The paraelectric states persist down to zero temperature for small~$\left( (\Delta^2/v_0^\perp ) / ( k_B T_c^0 ) \lesssim 0.01 \right)$  and moderate disorder~$\left( 0.01 \lesssim (\Delta^2/v_0^\perp ) / ( k_B T_c^0 ) \lesssim 0.02 \right)$. 
For small disorder, we now show that the non-polar states have an essential singularity: 
for $T=0$ and $p=0$, we solve in the classical limit the Euler-Lagrange equations~(\ref{eq:euler-lagrange}) for the zone-center transverse optic mode frequency,
\begin{align}
\label{eq:omega_DIS}
M ( \Omega_0^\perp )^2 &= \frac{ \left( 3 \gamma/2 \right) \left( A_0 / v_0^\perp \right) ( \Delta^2/v_0^\perp) }{ B^2 Q^2 } \nonumber \\ 
&
\times 
W_0\left[ \frac{4 B^4 Q^4 {v_0^\perp} e^{-1} }{ \left( 3 \gamma/2 \right) \left( A_0 / v_0^\perp \right) ( \Delta^2/v_0^\perp) }  e^{-2 B^2 Q^2 \frac{k_B T_c^0}{( \Delta^2/v_0^\perp) }}  \right].
\end{align}
Here, $W_0[z]$ is the zeroth branch of the Lambert function;~\cite{Corless1996a}
$A_0 / v_0^\perp \equiv (1/N) \sum_{\bm q} \left( v_0^\perp - v_{\bm q} \right)^{-1} = \left( (4\pi/\sqrt{3}) B Q \right)^{-1} $, 
$B^2 = ( \zeta a^2 )/( 4\pi /3)$,
$k_B T_c^0 = \left( v_0^\perp - \kappa \right) / \left(3 \gamma A_0 / v_0^\perp \right)$, and $Q$ is a wavevector cut-off.
Using that for $W_0[z] \simeq z + \mathcal{O}(z^2) $ for $z \to 0$, we obtain that in the limit of small disorder, 
\begin{equation}
\label{eq:essential_singularity}
M ( \Omega_0^\perp )^2 \simeq \left( 4 B^2Q^2 v_0^\perp e^{-1}  \right) \,  e^{-2 B^2 Q^2 \frac{k_B T_c^0}{( \Delta^2/v_0^\perp) }},
\end{equation}
which is an essential singularity. 
This points to the necessary occurrence of non-polar states with a stability region that
extends to zero temperature. 
First-order phase transitions are expected, nonetheless,
as the paraelectric state is close to a saddle point in the free energy 
for small disorder and well below $T_1$, as shown in
Fig.~\ref{fig:omega_p_small}. 

\begin{figure*}[htp!]
\includegraphics[scale=0.75]{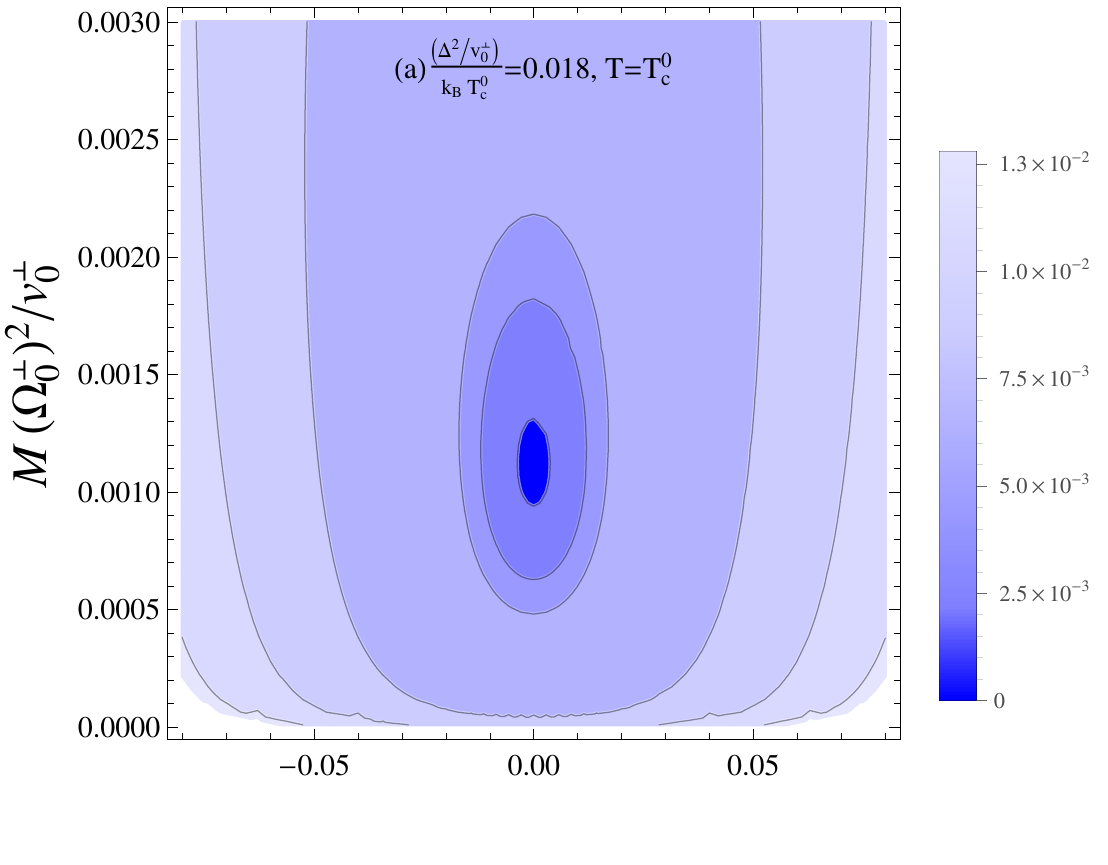}
\includegraphics[scale=0.75]{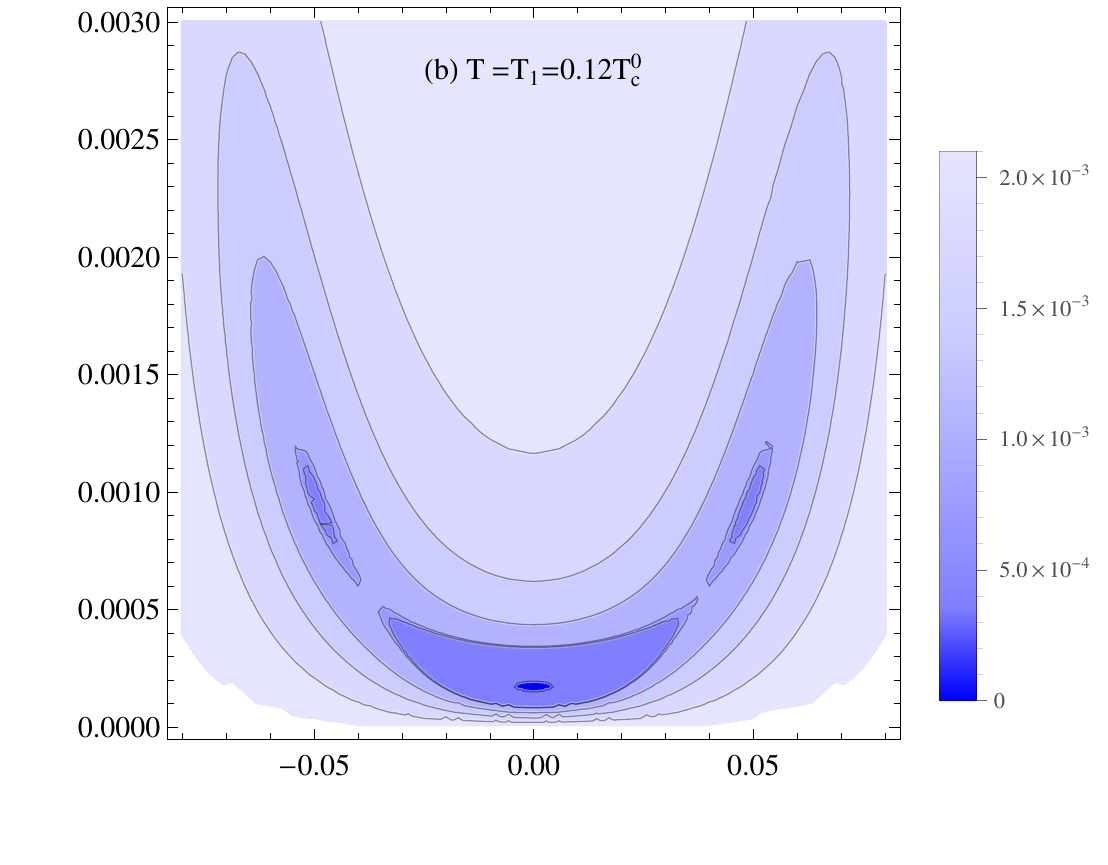}
\includegraphics[scale=0.75]{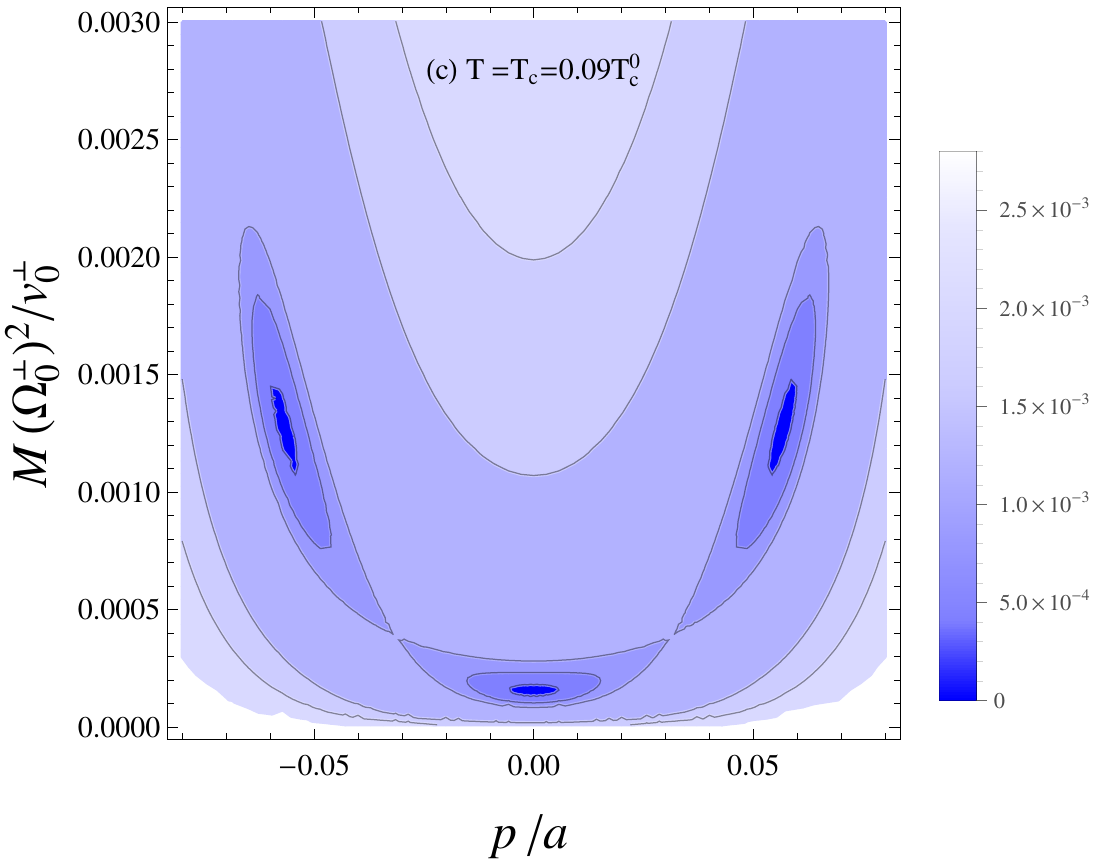}
\includegraphics[scale=0.75]{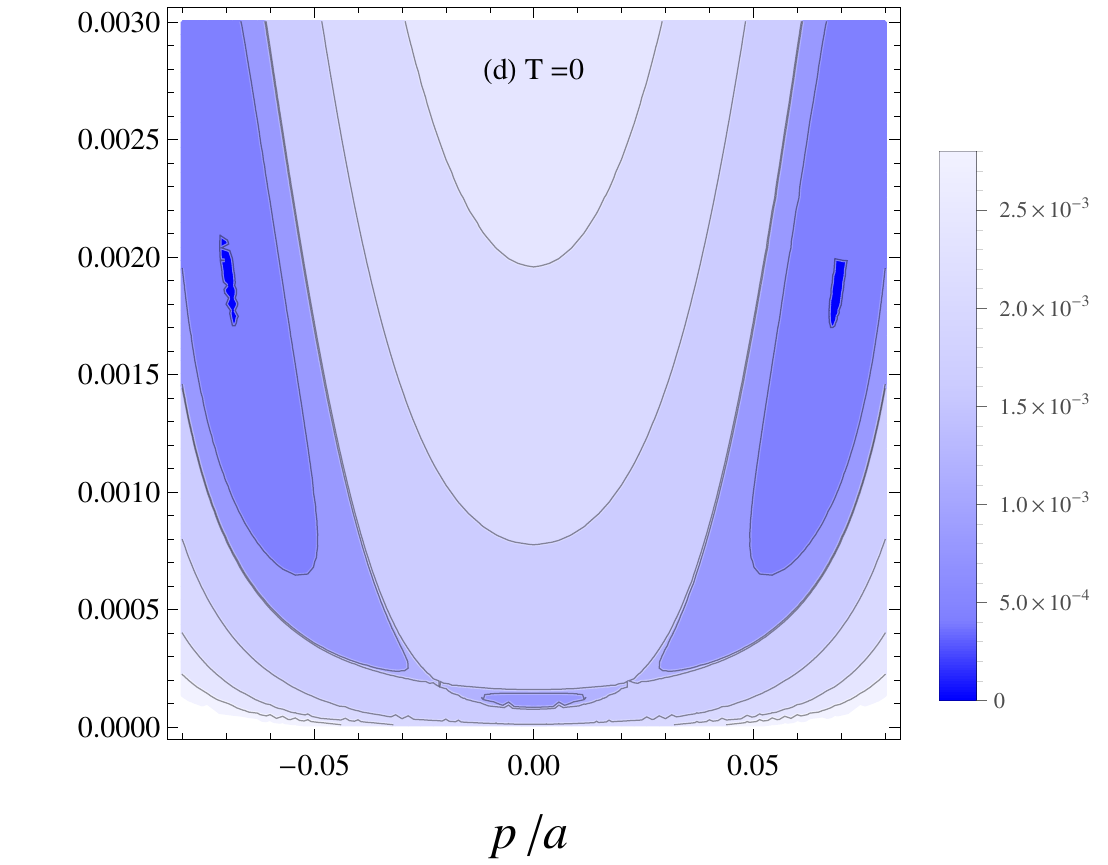}
\caption{(Color online)  Contour plots of the free energy for moderate disorder without applied electric fields. 
(a) For high temperatures there is paraelectric global minimum. (b) Below a superheating temperature $T_1$, ferroelectric states with 
opposite polarization appear as local minima. (c) 
At the temperature $T_c$, the polar and non-polar states coexist. (d) 
At $T=0$, the paraelectric state is metastable while 
the ferroelectric states are stable. 
Free energies are measured with
respect to that of the global minimum in units of $k_B T_c^0$. 
Here, $(\Delta^2/v_0^\perp)/(k_B T_c^0)=0.018$.}
\label{fig:free_energy}
\end{figure*}

\begin{figure}[h!]
\includegraphics[scale=0.75]{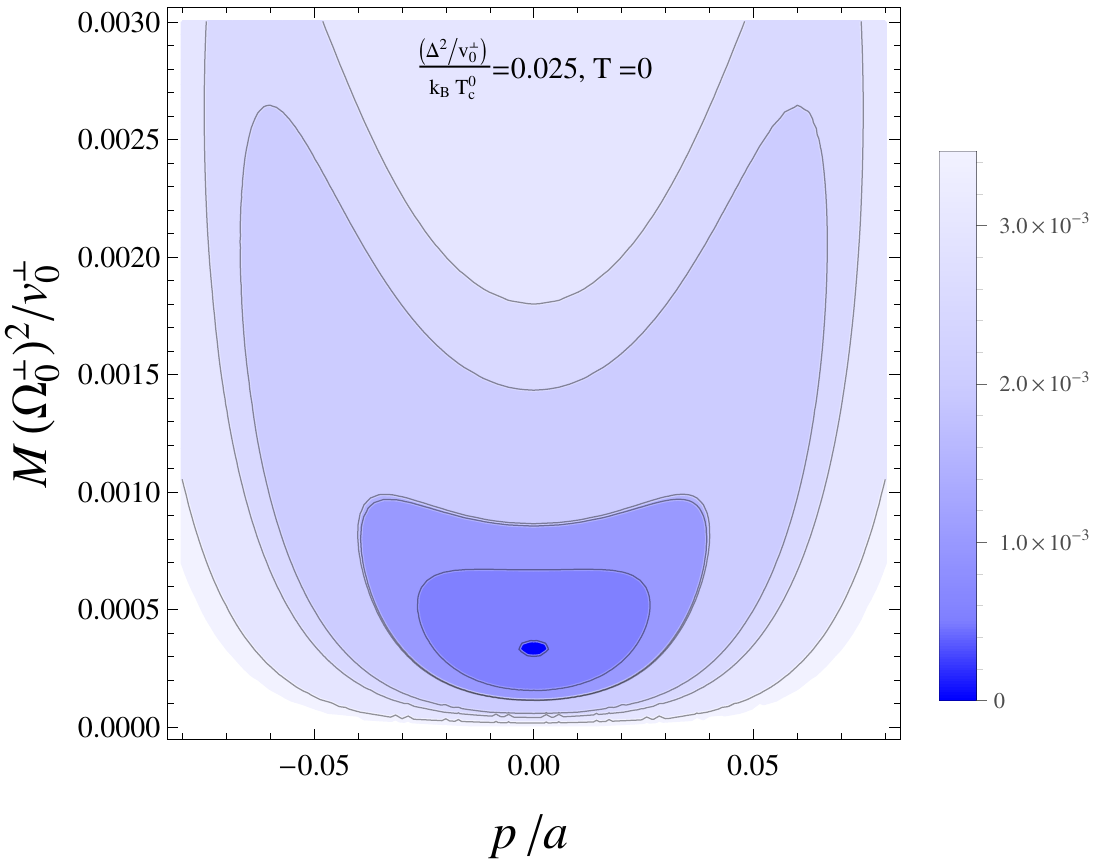}
\caption{(Color online) Contour plot of the free energy for large disorder. 
No ferroelectric states appear and 
there is only a global paraelectric minimum at all temperatures. 
Free energies are measured with respect to that of the global minimum in units of $k_B T_c^0$. }
\label{fig:free_energy_large_disorder}
\end{figure}

\begin{figure}[h!]
  \includegraphics[scale=0.65]{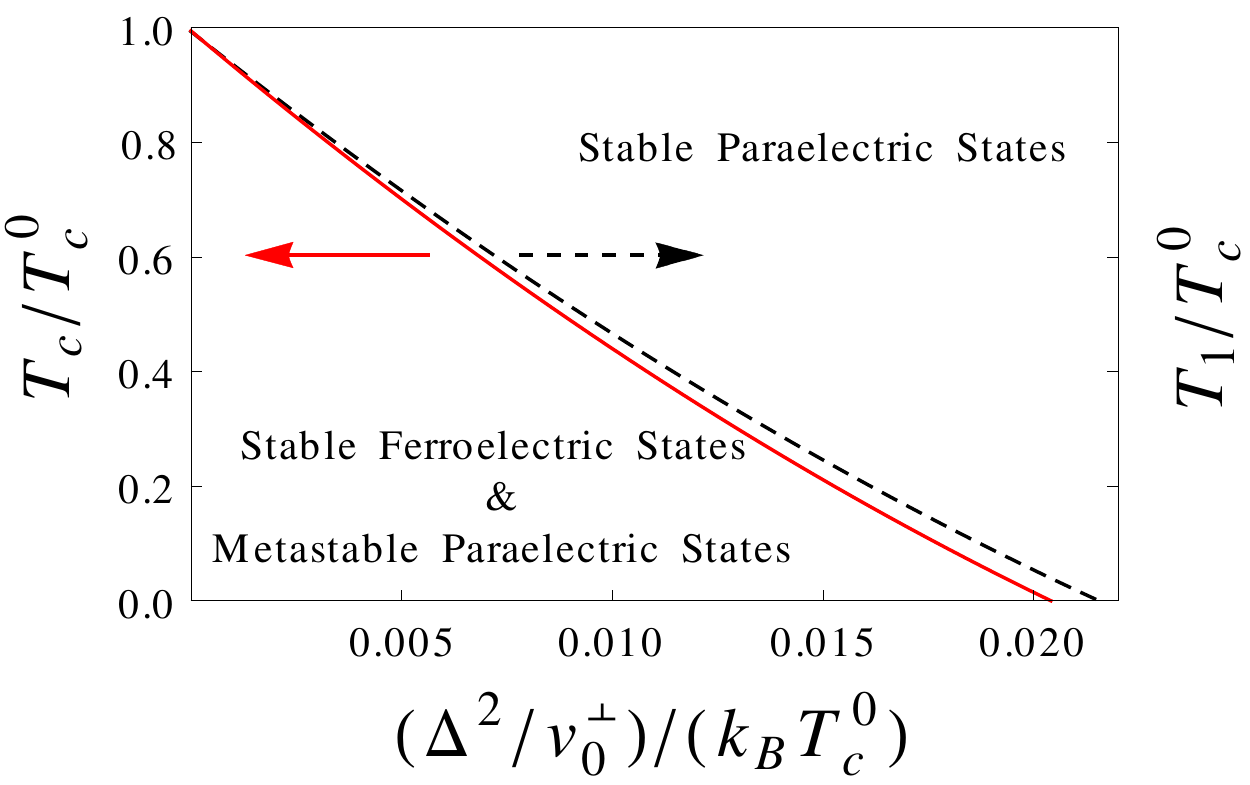}
  \caption{(Color online) 
            Disorder dependence of the coexisting and superheating temperatures $T_c$ and $T_1$, respectively.}
\label{fig:phase_diagram}
\end{figure}

\begin{figure}[h!]
  \begin{center}
      {\includegraphics[scale=0.6]{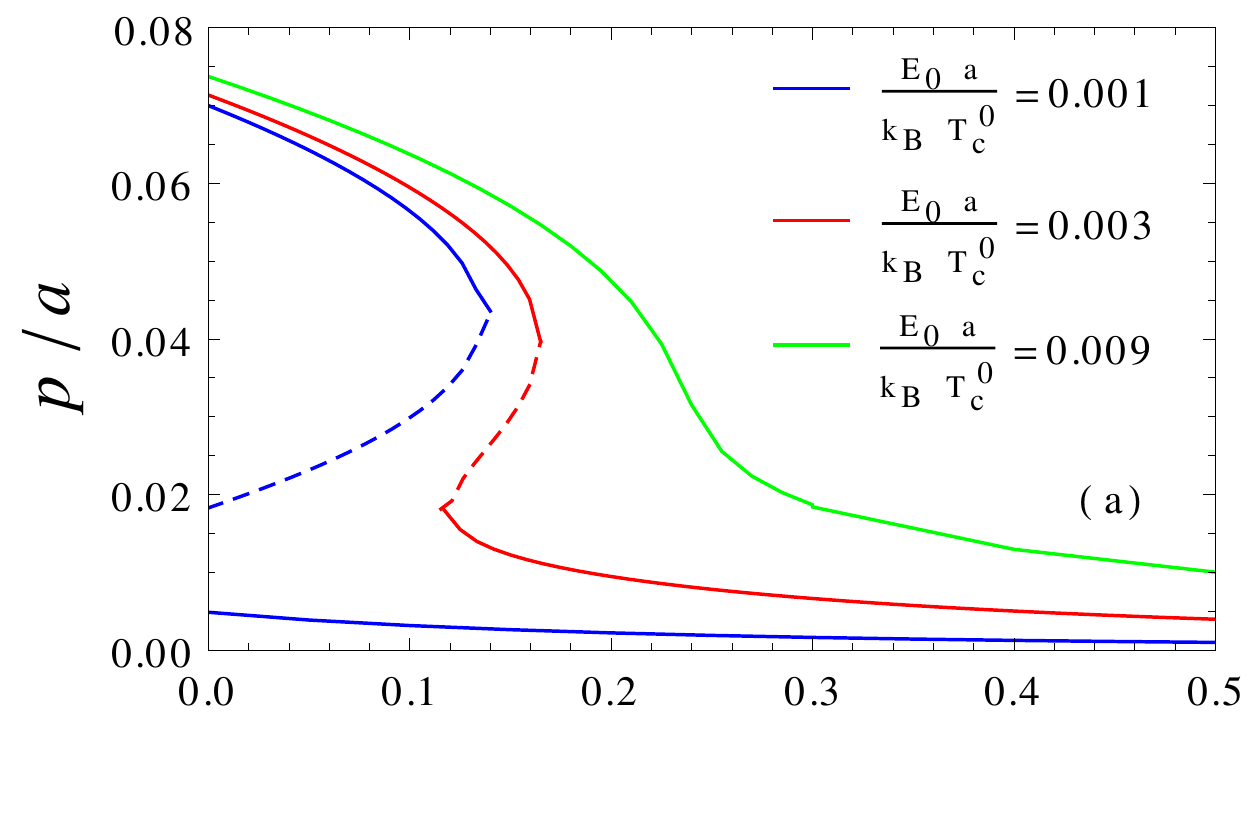}} \\ \vspace{-0.5cm}
      {\includegraphics[scale=0.6]{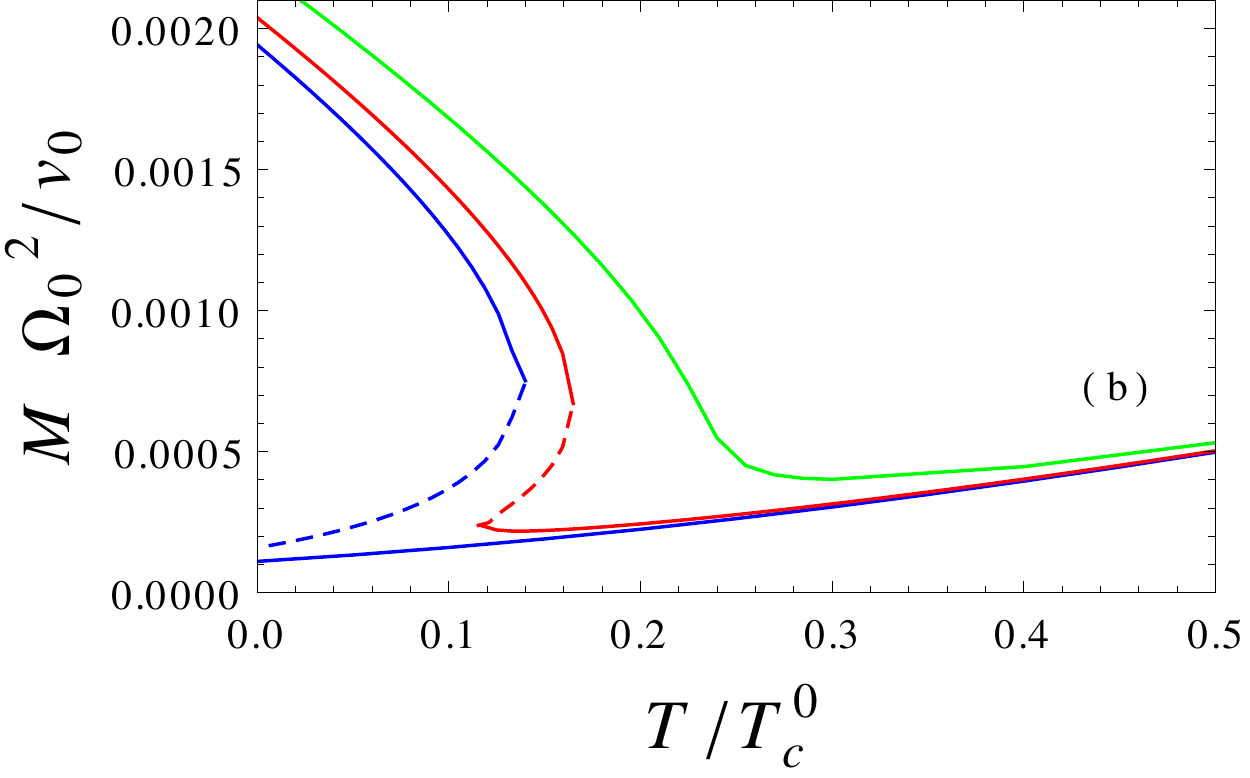}}
   \end{center}\vspace{-0.65cm}
\caption{(Color online) Temperature dependence of (a) the order parameter $p$ and (b) the zone-center soft mode frequency $\Omega_0^\perp$ for
several applied field strengths.
Solid lines correspond to stable and metastable states.
Dashed lines correspond to a saddle points in the free energy. Here, $(\Delta^2/v_0^\perp)/(k_B T_c^0)=0.018$.}
\label{fig:omega_p_with_field}
\end{figure}

\begin{figure*}[htp!]
\includegraphics[scale=0.75]{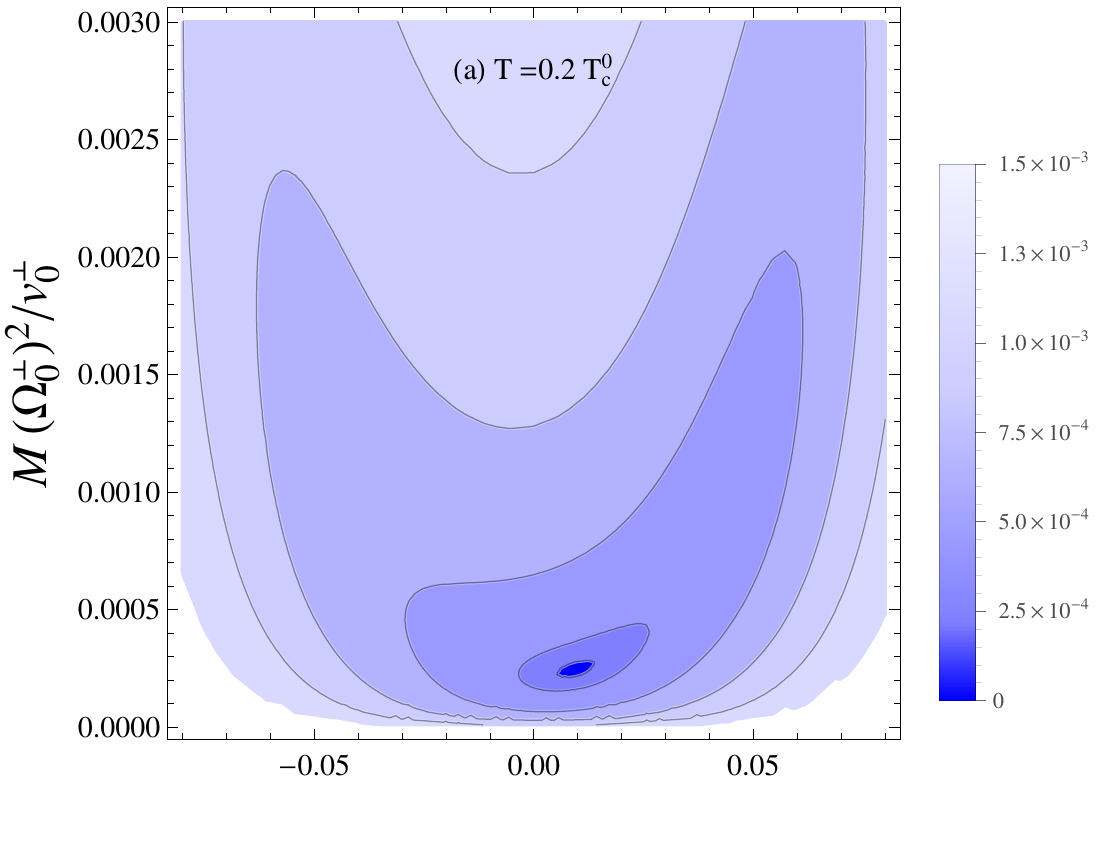}
\includegraphics[scale=0.75]{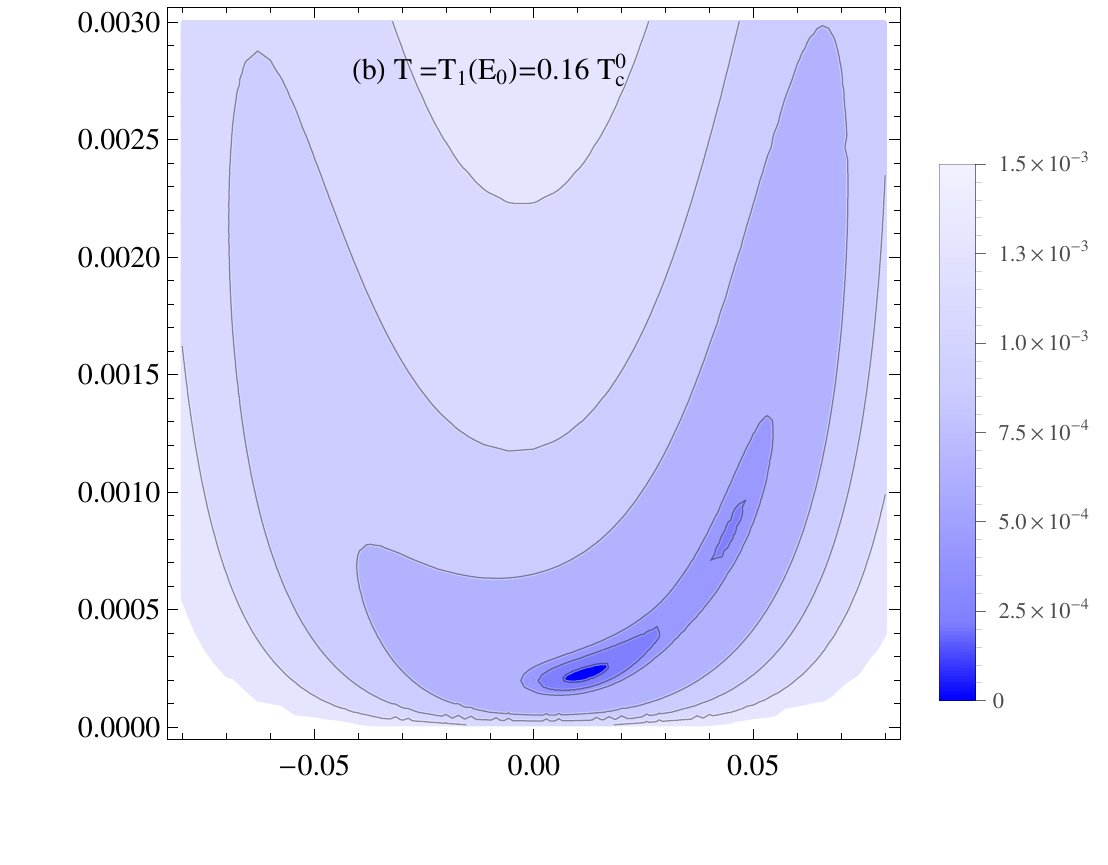}
\includegraphics[scale=0.75]{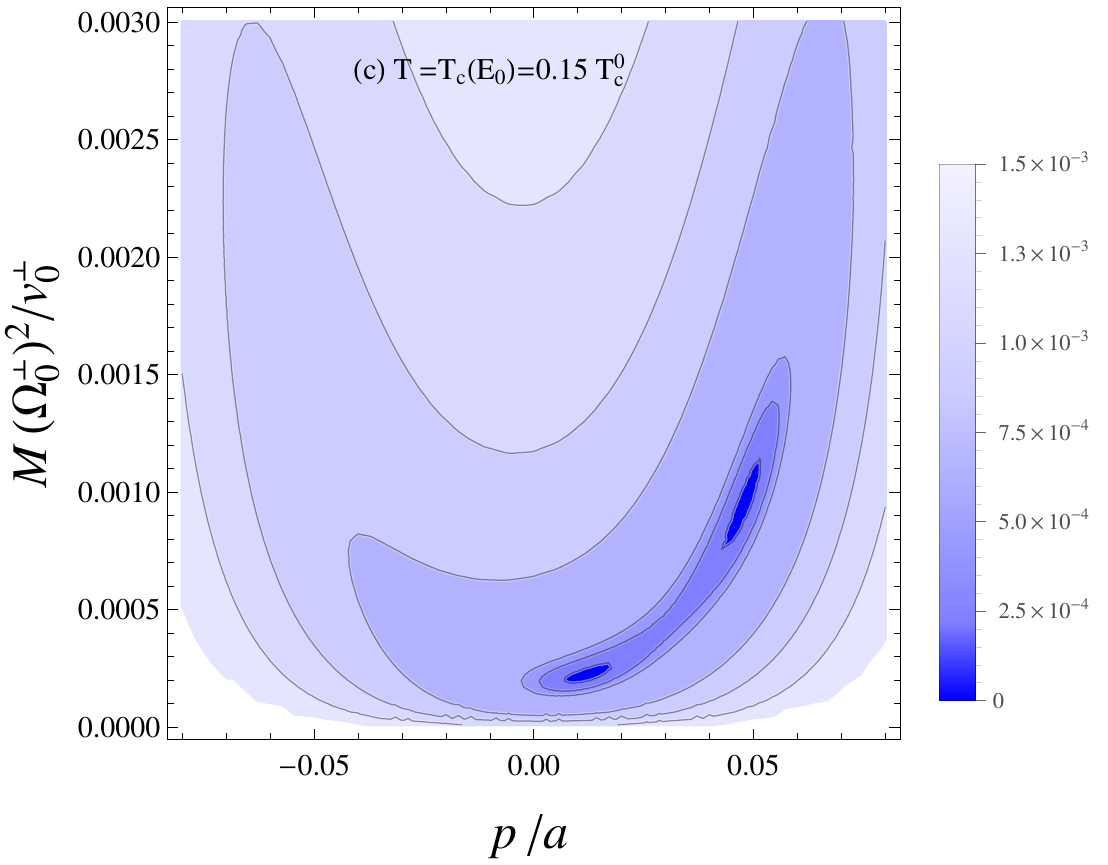}
\includegraphics[scale=0.75]{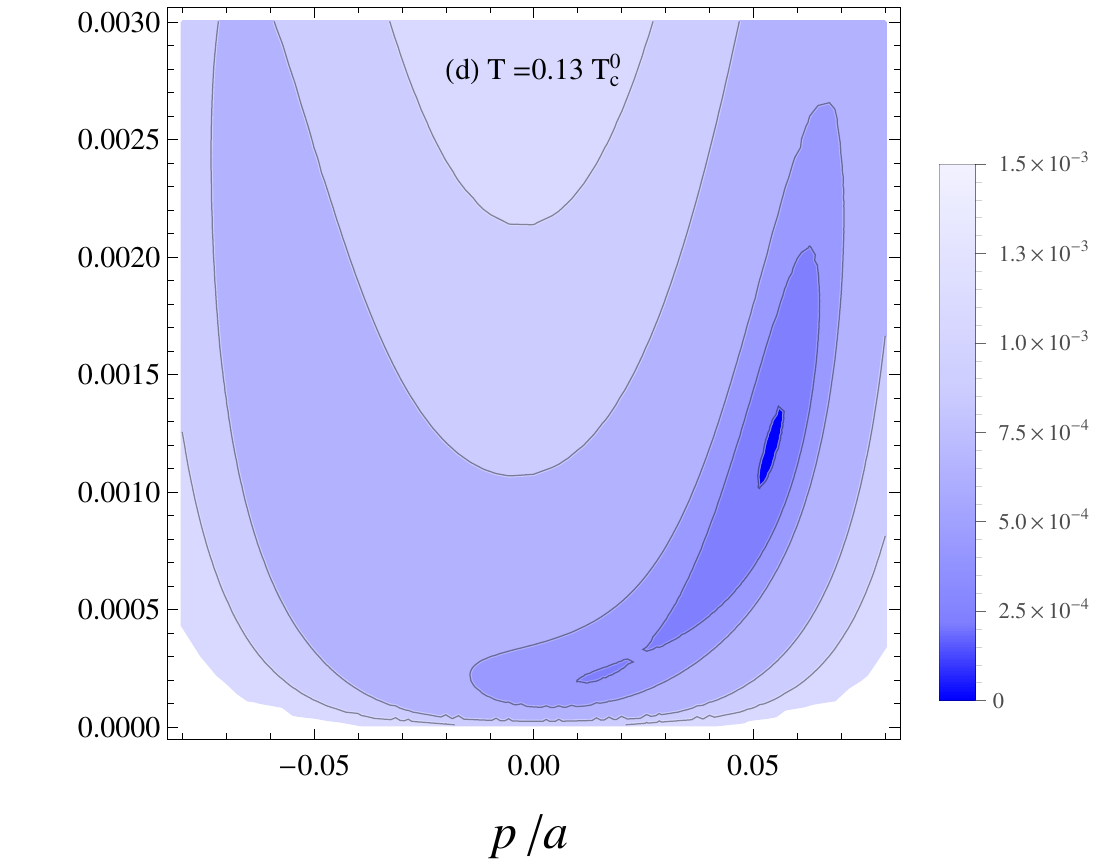}
\includegraphics[scale=0.75]{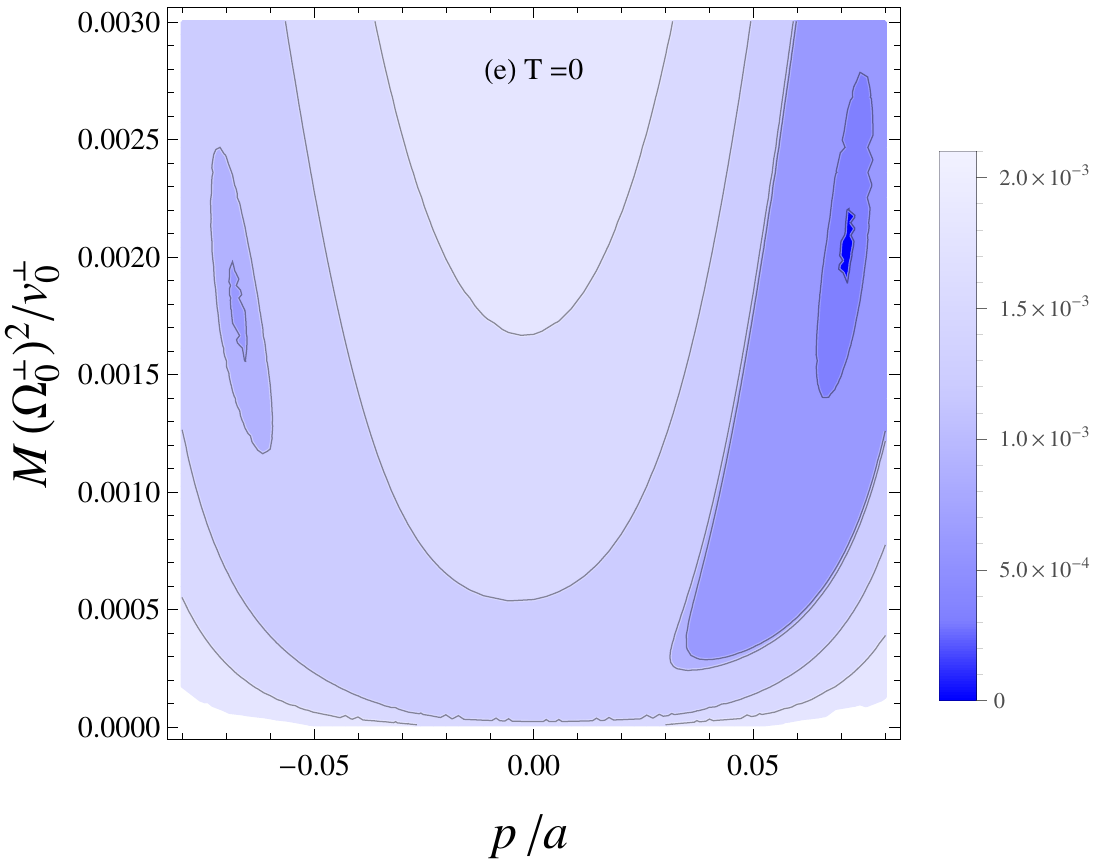}
\caption{(Color online) Contour plots of the free energy for moderate disorder and moderate applied electric fields. 
(a) For high temperatures there is a global paraelectric minimum with a polarization parallel 
to $E_0$. (b) At a temperature $T_1(E_0)$ a ferroelectric metastable state appears
with a polarization parallel to $E_0$. (c) 
At $T_c(E_0)$, the paraelectric and ferroelectric states coexist. 
(d) Below $T_c(E_0)$, the paraelectric state becomes a local minimum while the 
ferroelectric state is now stable.  
(e) For low temperatures, a new local ferroelectric minimum appears in the free energy 
with polarization opposite to $E_0$. 
Free energies are measured with
respect to global minimum in units of $k_B T_c^0$.
Here, $(\Delta^2/v_0^\perp)/(k_B T_c^0)=0.018$ and
$ E_0 a / ( k_B T_c^0 ) = 0.003 $. }
\label{fig:free_energy_with_field}
\end{figure*}

\begin{figure}[htp!]
\begin{centering}
\includegraphics[scale=0.6]{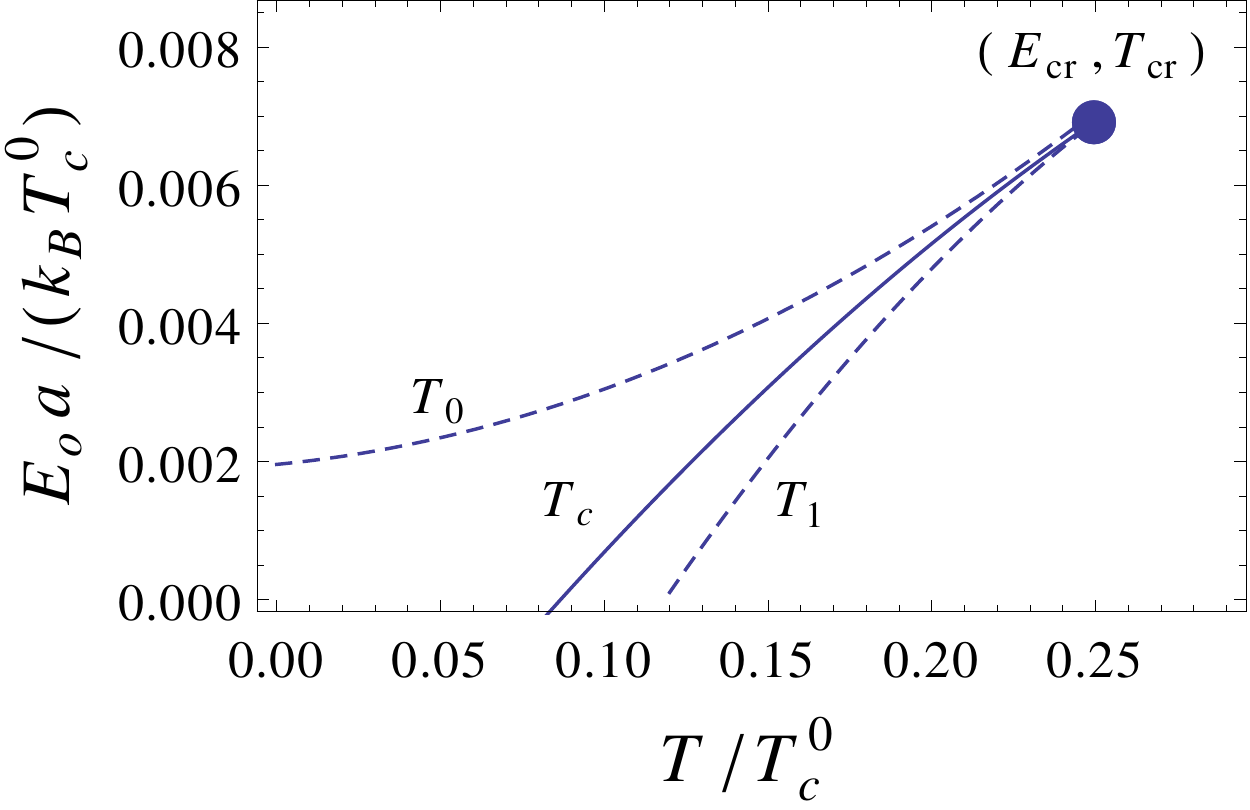}
\caption{(Color online) Calculated $E-T$ phase diagram for moderate disorder. 
$T_c$ is the coexisting line ending at the critical point $(E_{cr},T_{cr})$.
$ T_0$ and $ T_1 $ are spinodal curves that indicate the end of the stability
region of the paraelectric and ferroelectric phases, respectively.
Here, $(\Delta^2/v_0^\perp)/(k_B T_c^0)=0.018$. }
\label{fig:spinodal}
\end{centering}
\end{figure}

Fig.~\ref{fig:free_energy} shows the free energy landscape~($ p, \Omega_0^\perp $) for several temperatures.
For $T>T_1$ the paraelectric phase is stable. 
At $T_1$, ordered ferroelectric states appear as local minima. 
At a temperature $T_c < T_1 $ there is a coexistence region of disordered and ordered states. For $T < T_c$, the ordered states are stable and the disordered state is metastable down to $T=0$. Two saddle points with opposite polarization
appear at $T_1$. The temperature dependence of these saddle points is shown by the dashed line in Figs.~\ref{fig:omega_p_small}~(a)-(b).

The temperature dependence of the correlation length $\xi$ 
is shown in the inset of Fig.~\ref{fig:omega_p_small}~(b). For finite disorder, 
$\xi$ remains finite at all temperatures in the disordered states, though
it becomes very large for weak disorder 
as a consequence of the essential singularity~(\ref{eq:essential_singularity}).
For the ferroelectric states, $\xi$ grows as it approaches the superheating temperature $T_1$.

Figure~\ref{fig:phase_diagram} shows the temperature-disorder phase diagram.
Paraelectric metastable states extend from the the coexisting temperature $T_c$ down to zero temperature
for small and moderate disorder. 
First-order phase transitions are expected in the small disorder regime since 
the saddle points in the free energy are very close to the metastable paraelectric states at low temperatures,
as stated above. 
For intermediate disorder, 
no transition occurs for $T \to T_c$: the saddle and paraelectric points remain 
well separated~(see Fig.~\ref{fig:omega_p_small})
Nucleation of polar domains may occur within the paraelectric phase 
as stable ferroelectric states are present in the free energy.~\cite{Littlewood1986a}
For large disorder~$\left( (\Delta^2/v_0^\perp)/ ( k_B T_c^0 ) \gtrsim 0.02 \right)$, there is only a global paraelectric minimum in the 
free energy, as shown in Fig.~\ref{fig:free_energy_large_disorder}.

\subsection{Applied Electric Field, $E_0>0$.}

Figures~\ref{fig:omega_p_with_field}~(a)-(b) show 
the temperature dependence of the order parameter and the zone-center transverse optic frequency obtained from Eq.~(\ref{eq:euler-lagrange}) 
for finite applied fields and moderate compositional disorder. For clarity, only the states with polarization parallel to $E_0$ are shown.
For small field strengths~($ E_0 a / ( k_B T_c^0 ) \lesssim 0.002 $), the paraelectric states acquires a tiny polarization parallel to $E_0$
and persists down to zero temperature. The stable ferroelectric state does not change significantly from that without applied fields.
For moderate field strengths~($ 0.002 \lesssim E_0 a / ( k_B T_c^0 ) \lesssim 0.007 $),
first-order phase transitions are induced as the
the paraelectric states merges with the saddle point in the free energy. 
Upon increasing $E_0$, the transition smears out
for fields greater than a critical field $E_{cr}$~($ E_{cr} a / ( k_B T_c^0 ) \simeq  0.007 $), 
thus revealing a critical point.
The free energy landscape for moderate disorder is shown in 
Fig.~\ref{fig:free_energy_with_field}.

Figure~\ref{fig:spinodal} shows the $E-T$ phase diagram for moderate disorder. 
The coexistence line $T_c$ ends at the critical point $( E_{cr}, T_{cr} )$ 
above which the transition is smeared out. 
The spinodal curves $T_0$ and $T_1$ indicate 
the end of the stability region of the paraelectric and 
ferroelectric phases, respectively.
As opposed to pure ferroelectrics 
where it is observed that the spinodal curve $T_0$ of the paraelectric phase 
is close to the coexistence line and crosses the $T$-axis,~\cite{Meyerhofer1958a}
$T_0$ extends to zero temperature and does not
cross the abscissa  for finite disorder. 

\section{Comparison to Experiments}
\label{sec:comparison}

We compare our model with experiments in PMN and PMN-PT.

\subsection{No applied electric field, $E_0=0$}

In the absence of applied electric fields and for moderate compositional disorder,
our model shows that there is no symmetry-breaking down to low temperatures, 
as observed in PMN.~\cite{Bonneau1989a, deNathan1991a}
This disordered state is metastable and is not
that of a simple paraelectric, as their structure factors
are distinct,~see Eq.~(\ref{eq:Sq}).

We now estimate the Curie-Weiss temperature $T_{CW}$ and Curie-Weiss constant $C_{CW}$ for PMN from our model.  
By calculating the inverse dielectric susceptibility of our model,
$\chi^{-1} = M \left( \Omega_0^\perp \right)^2/ v_0^\perp$ 
for $ ( \Delta^2/v_0^\perp ) / ( k_B T_c^0 ) = 0.014 $~(see Sec.\ref{sec:applied_field}), 
we estimate by linear extrapolation that $T_{CW} \simeq 300\,$K and that $C_{CW} \simeq 5 \times 10^5 \,$K~($T_c^0 \simeq 720\,$K for PT~\cite{Glass1970a}).
This $T_{CW}$ is lower than the observed value~($ \simeq 400\,$K)~\cite{Viehland1992a} 
but it is consistent with a temperature higher than $T_c$.
The calculated $C_{CW}$ is slightly higher than that of experiments~($ \simeq 1.2  \times 10^5 \,$K ).~\cite{Viehland1992a} 
As opposed to $T_c$ and $T_1$, we find that $T_{CW}$ does not correspond to any special temperature in the free energy
of our model: it is simply the onset temperature of critical fluctuations of polarization
of a paraelectric state which do not undergo a phase transition. 

Well defined ferroelectric transitions are observed in 
PMN-PT with sufficiently large PT content~($\gtrsim 30\%$).~\cite{Shuvaeva2005a} 
This is consistent with our model as 
transitions to polar stable states are expected 
for small disorder~$ \left( ( \Delta^2 / v_0^\perp ) / \left( k_B T_c^0 \right) \lesssim 0.010 \right)$.
From Fig.~\ref{fig:phase_diagram}, we find that $T_c / T_c^0 \simeq 0.48$~(or $T_c \simeq 350\,$K) 
for $\left( \Delta^2 / v_0^\perp \right) / \left( k_B T_c^0 \right) = 0.010$. This is slightly lower but consistent with  the observed transition temperature of about $400\,$K in 
PMN-PT with $30\%$PT.

\subsection{Applied electric field, $E_0>0$}
\label{sec:applied_field}

In comparing to experiments in PMN in the presence of applied fields, 
we must distinguish between the observed behavior
of the skin and that of the bulk. 
Since the skin is macroscopically large~(a few tens of micrometers),~\cite{Stock2007a, Conlon2004a}
we compare our model to the skin and the bulk separately.

We first compare to the skin of PMN.
Our model
shows there is a field induced first-order phase transition for 
applied fields that are large enough to cross the stability limit 
of the paraelectric phase, see Fig.~(\ref{fig:spinodal}). Upon increasing the applied field,
the transition smears out above a critical field $E_{cr}$,
as observed in PMN.~\cite{Kutnjak2006a} 
We can estimate $E_{cr}$ from our model:
by fitting the superheating temperature $T_1$
to that of field-cooled-zero-field-heating experiments in PMN~($T_1 \simeq 210\,$K),~\cite{Schmidt1980a, Zhao2007a}
we find that the corresponding disorder strength is of about 
$ ( \Delta^2/v_0^\perp ) / ( k_B T_c^0 ) = 0.014 $,
according to Fig.~\ref{fig:phase_diagram}. For this disorder strength, 
the transition is smeared at about $ E_{cr} a / (k_B T_c^0) = 0.004$, which corresponds to
about $5\,$kV/cm.~\cite{footnote2} 
This is close to the observed value of about $4\,$kV/cm.~\cite{Kutnjak2006a} 
Morphotropic phase boundaries between structurally distinct ferroelectric phases are observed in the skin
of PMN-PT for PT concentrations of about $30-35\%$.~\cite{Chois1996a, Noblanc1996a, Kelly1997a, Park1997a}
We cannot discuss this effect within our approximation as we do not consider cubic symmetry. 

We now compare to the bulk of PMN.
Bulk PMN does not go through 
any macroscopic structural phase transition under applied fields,~\cite{Stock2007a}  which is in disagreement with our model.
We believe this discrepancy arises because we ignore coupling to 
strain fields.
Despite there is no observation of global broken symmetries, 
neutron scattering experiments
reveal a smooth peak in  the diffuse scattering 
upon application of an electric field
precisely at about $T_c \simeq 200\,$K.~\cite{Stock2007a} 
This is suggestive of clamping effects~\cite{Cowley1996a} 
for which there are two possible scenarios:
(i) the bulk is clamped and remains in the metastable disordered state
while the skin can relax to access the stable states with spontaneous polarization; or (ii)
clamping effects are such that the disordered phase is stable down to
$T=0$ and the skin is in a metastable ordered state.
Previous theoretical studies consider coupling to acoustic modes but do
not address this point.~\cite{Burton2006a, Tinte2006a} 

\section{Conclusions}
\label{sec:Conclusions}

We have studied the effects of quenched 
random fields in a simple displacive model for ferroelectrics using a variational method.
We show that for small and moderate disorder there are metastable paraelectric states in the free energy
with a stability region that extends to zero temperature.
For small disorder, these states exhibit an essential singularity in their free energies. 
Ferroelectric states appear as local minima below a superheating temperature $T_1$
and above a coexisting temperature $T_c$. 
Below $T_c$, the ferroelectric states become stable.  
No global symmetry breaking occurs for moderate disorder
as the saddle points and disordered states remain well separated down
to zero temperature.  
First-order phase transitions are induced
for electric fields large enough
to cross the stability limit of the paraelectric phase.
These paraelectric and ferroelectric 
states have distinct structure factors from those
of conventional ferroelectrics. 

Based on our results, 
we present our view of the static thermodynamic behavior of heterovalent relaxors.
Pure relaxors such as PMN and PZN 
are in a stable paraelectric 
state for temperatures above $T_c$;
below $T_c$, they remain in a metastable disordered state 
with a stability region that extends down to zero temperature. 
Nucleation of local polar domains 
within the non-polar phase may occur as there are stable ferroelectric states in the free energy.
First-order transitions are induced for applied fields
large enough to cross the stability 
region of the metastable paraelectric phase. 
Upon increasing the applied field, the coexisting region approaches 
a critical point.  
The paraelectric and ferroelectric states are not those 
of conventional ferroelectrics as their structure factors differ. 
$T_{CW}~(\simeq T^*)$ is the onset temperature of critical fluctuations of polarization
of a paraelectric state which does not undergo a phase transition.
It does not correspond to a special temperature in the free energy within
the variational solution of our model.
Ferroelectric transitions occur in 
PMN-PT and PZN-PT with sufficiently 
large PT content as they fall in the weak disorder regime where the metastable disordered
state is close to an instability point. 

We suggest that clamping effects are responsible for
the lack of macroscopic symmetry breaking 
in the bulk of PMN and PZN upon application of 
electric fields. Extensions of this model to incorporate cubic symmetries and coupling
to acoustic phonons are needed to validate or refute
this point. 

\section{Acknowledgments}
GGGV and PBL acknowledge useful discussions with 
Stephen Streiffer and Ray Osborn. 
Work at Argonne is supported by the U.S. Department of Energy, 
Office of Basic Energy Sciences under contract no. DE-AC02-06CH11357.
Work at UC Riverside supported by the UC Lab Fee Program 09-LR-01-118286-HELF. 
GGGV and CMV wish to thank other principal investigators with whom the UC grant was issued: Frances Hellman, Albert Migliori and Alexandra Navrotsky.


\begin{thebibliography}{99}
\bibitem{footnote2}
Simple dimensional analysis shows $E_0$ must be replaced by $E_0 e^*$ in our model
to obtain the experimental electric field.
\end{thebibliography}
\end{document}